# Royal Observatory of Belgium Gravitational balance

**Michel van Ruymbeke**, Sébastien Naslin, Ping Zhu, Francis Renders & Jean-Philippe Noël

Royal Observatory of Belgium, Av.Circulaire, 3, 1180 Brussels, Belgium

labvruy@oma.be

**Abstract:**

An astatized symmetrical vertical pendulum is monitoring torque $\Gamma(M)$ resulting of gravitational attractions exerted by two external masses M moving up down. Local gravity field g produce the main pendulum restoring torque combined with a very weak variable torque $\Gamma(c)$ induced by rotation of watch needles fixed to the pendulum. Transfer of fundamental units to calibrate the $\Gamma(c)$ torques is obtained by a reference torque $\Gamma(\mu)$ resulting of precise displacements of a well known mass µ. We permanently monitored ratio between the gravitational effect $\Gamma(M)$ and calibrated $\Gamma(c)$ to determine G. Position of the pendulum is measured with a capacitive bridge. Bias voltages sent to two electrodes set-up at the bottom of the pendulum allows to feedback pendulum with a controlled electrostatic torque. We discuss potential interest of our prototype to design a multi pendulum system to check systematic effects for different geometries and various kinds of materials.

*Keywords:* gravitational constant, metrology, pendulum

## Introduction.

Determination of the G-gravitational constant constitutes a challenge for the metrology. CODATA 2006 announced: G = 6,67428(67) ± 0,0006.
However new determinations *[Parks, H.& Faller, J.E., 2010] [Liang-Cheng Tu & al.2010]* confirm that the operating mode could induce different systematic effects. Purpose of our researches consists to overview possibility from tidal instrument expertise, to measure G in function of the local g with absolute precision better than 0.1ppm *[van Ruymbeke, M.,1979]*.

### 1. Properties of the Symmetric Vertical Pendulum G-svp design.

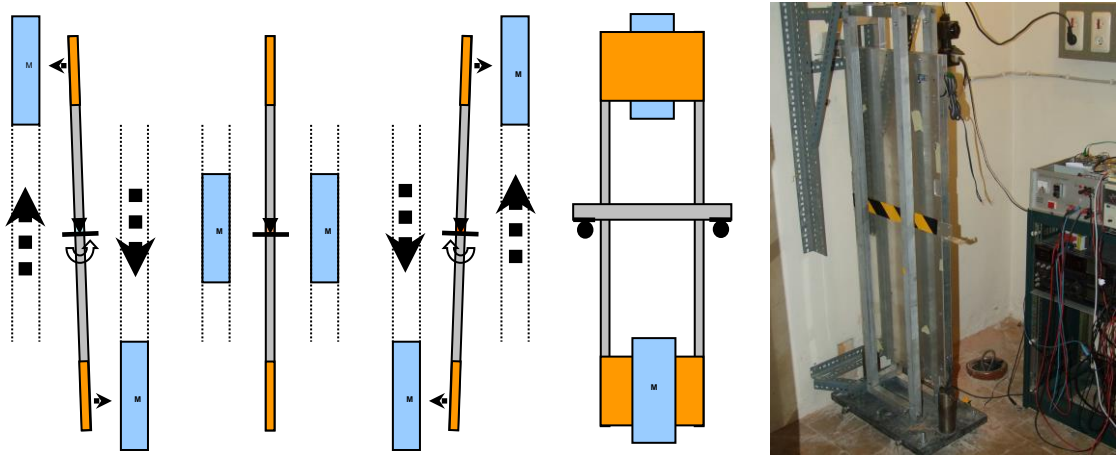

*Fig1: (left): Principle of G measurement. Gravitational field acting on the pendulum masses is modulated by the displacement of two aluminium masses. A bar fixed in the centre of the pendulum body, maintains rigidly the two lames used to fix the axis of rotation of the pendulum. The symmetric position of the pendulum between the two alternate positions of masses allows decreasing effect of positioning error. (right): View of the G-svp system.*

Design of G-svp micro force balance using the EDAS concept *[van Ruymbeke M &al.]* is dedicated to G measurement (*fig1*) answering to some priorities:
- Basic choice of pendulum approach *[Naslin, S., 2009]* referred to g with actions expressed in term of torques Γ with at equilibrium Σ Γ = 0.
-Astatization obtained with a horizontal axis of rotation just over centre of mass of pendulum allowing comparing very weak horizontal forces with larger vertical ones. Central positioning rejects largely translation mode of the micro seismic noise acting on the support.
- Geometry defining the gravitational torque for different positions of the moving parallelepiped masses with like a "cross" design to decrease the effect of lateral discrepancy.
- Minimisation of thermal, acoustic and magnetic influences induced by the motion of the attracting masses. Design of vertical attracting masses motion shows constant loading effect on the ground rejecting risk of pendulum tilt.
- Axis of rotation experimental selection of cutter knives Stanley N°1992 with their edges set perpendicular on 3mm in diameter drill rig cylinders.
- Use of electronics based on symmetric phase detector *[van Ruymbeke, M., 1980]* connected to capacitive transducer *[van Ruymbeke M, 1980]* with a limited gap between plates large enough for pendulum operating quite at critical damping just by air friction. It allows very high precision in the positioning without hysteresis.
-Application of bias signal using the bottom capacitance plate to induce electrostatic torques for determination of step response or to apply electrostatic feedback to keep constant the position of the pendulum mass allowing neglecting the inelasticity of the suspension. PWM signals on both sides decreases non linear coupling between electrostatic induction and instantaneous position of the mass vibrating under environmental influences.
- A well calibrated very stable modulated action should be applied to the pendulum simultaneously with gravitational attraction, but with a different periodicity. So it becomes possible to determine the amplitude of both signals by stacking on each period. Ratio between the two amplitudes is independent of the sensitivity of the pendulum. Stability of time modulation periodicities allows signal-to-noise improvement by stacking on very long time intervals.

## 3. Scale factor determination.
The displacement of a weight mg fixed to the pendulum body induces a variable torque only function of the horizontal displacement in front of the axis of rotation. A first system consists to put horizontally on the pendulum a quartz watch (*fig2,left*). The weights of the needles rotating induce a very stable sinusoidal torques Γ(c).

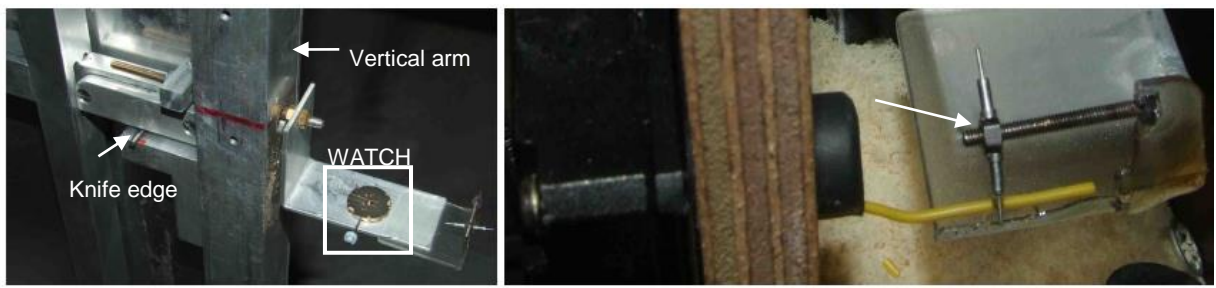

*Fig2: (left). Torques induction to the pendulum by the rotations of quartz watch fixed rigidly to the pendulum on an external support; (right) A second system consists to move a small nut with two axial terminations turning at constant speed on a screw with a pitch of 0.400 mm.*

We calibrate the clock torques Γ(c) with (*fig2,right*) a second torque Γ(μ) induced by a known mass of 195,0019 mg displaced with a constant speed of ~ 1μm/sec by step motor knocking regularly with a small arm parallel to x-axis Multi steps with high frequency allow rejecting friction effects.

The procedure consists (*fig3*) to register simultaneously signals of the two torques Γ(c) and Γ (μ) to determine their ratio. After such calibration procedure, only clock rests active with a very soft effect on the G-SVP balance inducing very constant calibrated micro-torque to determine G-svp scale factor.

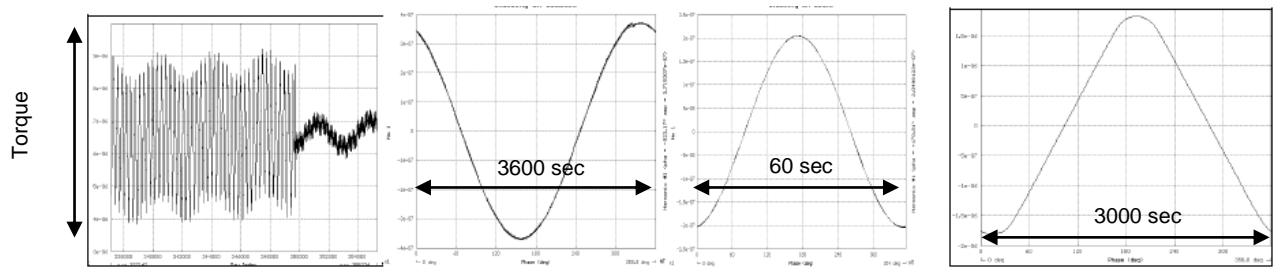

*fig3: (left) Example of the torques record with and without μ mass rotation; (centre) We obtain by stacking average of the signal modulation [van Ruymbeke M.,2003] an amplitude of 3.7153E-7Nm for one hour period and an amplitude of 2.0498E-7Nm for one minute; (right) The speed of μ mass rotations on its screw fixes the Γ(μ) modulation. Sensitivity is obtained by a stacking procedure on the period of 5 minutes corresponding to ±2x6 turns of the step motor (±2.4mm).*

## 4. The gravitational effects registered with G-SVP balance

We model gravitational response with the parameters selected for the pendulum reacting to the two masses motion (*fig4,left*). We extract with averaging by stacking *[van Ruymbeke M.,2003]* the signature of gravitational modulation after normalisation of real records by the sensitivity factors obtained with the calibration process described before (*fig4,right*).

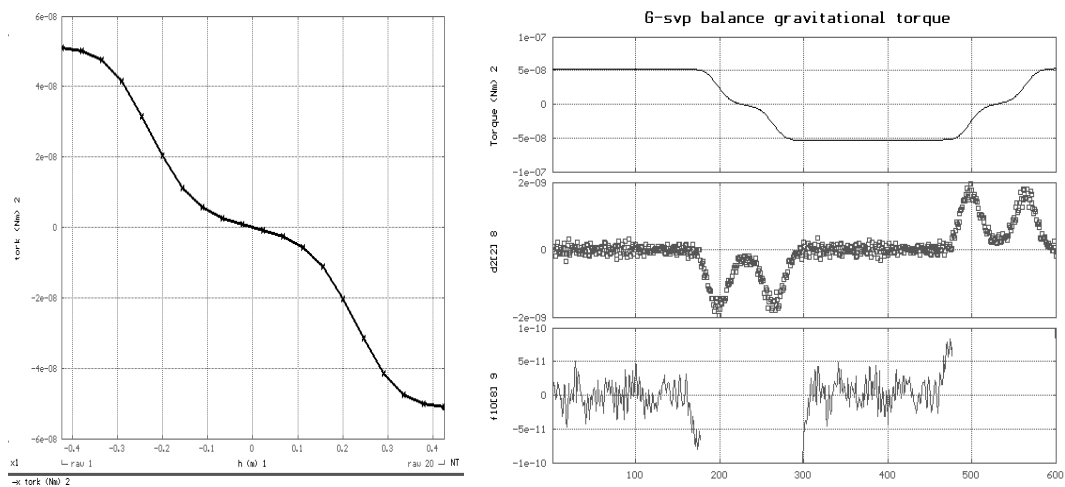

*fig4:(left): Theoretical model of the torque modulation expressed in function of the levels h of the masses comprised between +0.4m and – 0.4m.;(right) Channel 1 shows stacking on 10 minutes period of Gravitational effect. Channel 2 is the derivative of channel 1 stacking. Channel 3 is a zoom of filtered by sliding average of channel 2 showing noise amplitude corresponding to a dynamics of 1/1000 on gravitational effect.*

Comparison with theoretical signature confirms quality of result. A coherent result by comparison of admitted G constant value is achieved at 1% of accuracy. A first derivative shows a signal-to-noise ratio of about 0.1% on the stacking. By mathematical adjustment for longer series in well sited laboratory, we could have access to better level of confidence. If we replace masses selected for our prototype constituted of aluminium by copper plates, the amplitude of gravitation torques could be multiply by ten. So a dynamics better of 100 ppm becomes realistic.

## Conclusions

Our prototype simply made with aluminium is characterized by promising patterns. It operates without vacuum and shows a high rejection of ground vibration. The large enough support with lames fixing axis of rotation could work for heavy masses without inelastic problem. The pendulum is calibrated by a fully independent action simultaneously with gravitational measurements. Geometrical discrepancies are well rejected by a symmetrical approach. It could be justify improving our methods by a more careful setting of our balance.

With high density material, gravitational torques could be multiplied by ten. Pendulum set-up in underground laboratory at constant temperature and pressure could potentially increase the significance of results. We evaluate for the G-svp balance the signal-to-noise accessible to 100 ppm. Final goals could be to build a three parallel pendulums system with the two attracting masses passing through both of them to reach effective information about systematic errors figure. By comparison of individual output signals it could be possible to manage detection of all kind of effects influencing G-constant value. This could be a good system to experience gravitational physics.

## Acknowledgements

The authors are appreciative to the Royal Observatory of Belgium who provided the necessary support. We thank especially Eric Putz, André Somerhausen, François Beauducel, Eric de Kerchove and Geneviève Tuts who participate to the development of the EDAS concept applied for this project. Funding by Ing G.Berthault was determinant in the achievement of this research.